\documentclass[twocolumn]{article}

\usepackage{PRIMEarxiv}
\usepackage[utf8]{inputenc}
\usepackage[T1]{fontenc}
\usepackage{hyperref}
\usepackage{url}
\usepackage{amsfonts}
\usepackage{nicefrac}
\usepackage{microtype}
\usepackage{fancyhdr}
\usepackage{graphicx}
\usepackage{amsmath,amssymb}
\usepackage{algorithmic}
\usepackage{algorithm}
\usepackage{textcomp}
\usepackage{xcolor}
\usepackage{bm}
\usepackage{booktabs}
\usepackage{siunitx}
\usepackage{multirow}
\usepackage{adjustbox}
\usepackage[normalem]{ulem}
\usepackage{comment}
\providecommand{\Description}[1]{}

\graphicspath{{media/}}  

\title{Dependable Voice Command for Multi-UAV Swarms Using Small Language Models}

\author{
  Afshin Ameri$^{*}$, Luiz Giacomossi$^{*}$, Håkan Forsberg, Ivan Tomasic, and Baran C\"ur\"ukl\"u \\
  Mälardalen University \\
  Västerås, Sweden \\
  \texttt{\{afshin.ameri, luiz.giacomossi, hakan.forsberg, ivan.tomasic, baran.curuklu\}@mdu.se}
  \\
  \noindent\textit{*These authors contributed equally and share co-first authorship.}
}

\begin{document}

\twocolumn[
  \begin{@twocolumnfalse}
    \maketitle
    \begin{abstract}
        Supervisory control of multiple autonomous unmanned aerial vehicles (UAVs) imposes high visual and cognitive demands on human operators, particularly when communicating mission objectives and system constraints. Voice command provides a natural interaction modality that can reduce operator workload; however, its reliability is affected by environmental noise, speech variability, and recognition errors. This paper presents a dependable voice command framework for multi-UAV swarm supervision using Small Language Models (SLMs). The proposed architecture integrates an Automatic Speech Recognition (ASR) module with an SLM-based semantic processing layer to transform spoken operator commands into structured task definitions for downstream multi-UAV planning and execution. To evaluate robustness, an end-to-end assessment pipeline is developed using a Probabilistic Context-Free Grammar (PCFG) to generate synthetic supervisory commands, which are converted into audio samples with varying operator accents and environmental signal-to-noise ratios (SNR). The resulting speech signals are processed by the ASR module, and the SLM performs contextual interpretation and phonetic correction of transcriptions when required. The evaluation isolates the resilience of the voice command pipeline under acoustic degradation and speech variation, demonstrating reliable task generation for multi-UAV supervision in challenging operational conditions.
    \end{abstract}
    
    
    \vspace{1cm}
  \end{@twocolumnfalse}
]

\section{Introduction}
\label{sec:introduction}

Supervisory control of multiple autonomous unmanned aerial vehicles (UAVs) imposes high visual and cognitive demands on human operators \cite{hocraffer2017meta}. This is partly related to interacting with the system to monitor mission progress and to inform the system of mission parameters. While speech interaction provides a complementary communication modality, unregulated text-to-speech (TTS) mechanisms often induce excessive verbosity and cognitive overload \cite{pruitt2023informing, donmez2009auditory}. Unregulated speech input relies on Automatic Speech Recognition (ASR), which degrades under environmental noise and cross-cultural phonetic differences \cite{koenecke2020racial}. Standard implementations lack the structural constraints required for dependable task generation in multi-agent environments.


For system input, voice control introduces distinct operational hazards. Unregulated commands rely entirely on the transcription accuracy of the ASR engine. In multi-UAV environments, ASR degrades under non-stationary acoustic masking from rotors and cross-cultural phonetic variations among operators. Relying on raw ASR transcripts causes referential ambiguity which might lead to misclassification of target identities. Static speech-to-text pipelines execute unsafe commands when acoustic distortions alter the intended instruction.

To resolve transcription degradation, the input architecture must validate operator intent before execution. This architecture processes raw ASR outputs through a Small Language Model (SLM), which operates as a semantic parser. It translates dialect-heavy and noisy transcripts into task definitions (JSON). It can also adjust for phonetic similarities that might have affected the ASR (e.g., the word "zoom" might be transcribed as "soon" in ASR). This isolates the stochastic uncertainty of speech recognition from the deterministic task allocation logic of the drone swarm.

These design strategies shift the multimodal interface from reactive alerting to a regulated, bidirectional communication channel. The input policies applied to the audio input guarantee a safe task generation in degraded operational contexts.



    The contributions of this work are a semantic interpretation pipeline based on a Small Language Model (SLM) that enables the extraction of operator intent from culturally diverse speech inputs and their transformation into structured JSON-based task representations for downstream drone allocation. The pipeline further incorporates phonetic adaptation mechanisms to mitigate applicable ASR transcription errors. Additionally, this work introduces an end-to-end stochastic evaluation methodology for quantifying system dependability under controlled acoustic degradation conditions (SNR variations) and cross-cultural dialect variability.

\section{Related Work}
\label{sec:related_work}
In this section, we review multimodal interfaces in UAV swarm supervisory control, focusing on the use of auditory feedback and speech input, along with their limitations and challenges.
\subsection{Multimodal Interfaces in Supervisory Control}

Supervisory control of UAV swarms saturates the operator's visual channel \cite{hocraffer2017meta}. To mitigate visual overload, multimodal interfaces utilize bidirectional speech interaction. For system output, auditory feedback reduces attentional demand \cite{gonzalez2012psycho} and decreases reaction times to system deviations \cite{donmez2009auditory}. For system input, voice commands enable operators to direct multiple agents without disengaging visual attention from spatial displays \cite{park2023vehicle, contreras2020unmanned}. This interaction modality is particularly critical in Manned-Unmanned Teaming (MUM-T) architectures, where a human pilot must issue rapid tactical assignments to autonomous aircraft while actively managing the leader vehicle \cite{ricardo2023cooperative}.

However, unregulated speech interfaces introduce operational risks. On the output side, continuous auditory signals cause alert fatigue and channel saturation \cite{pruitt2023informing}. Empirical evaluations of remote vehicle operators demonstrate that the efficacy of auditory feedback is strictly bounded by operator workload. Dunn et al. \cite{dunn2022measured} showed that while auditory feedback improves decision-making under low workloads, it acts as an external stressor and degrades spatial orientation under high workloads. Uncalibrated auditory volume further deteriorates operator accuracy \cite{dunn2023effects}.

On the input side, voice control introduces referential ambiguity and relies heavily on the accuracy of Automatic Speech Recognition (ASR). Command execution fails when ASR misclassifies target identities or action parameters. Consequently, static speech interfaces are insufficient for multi-agent supervision. Auditory output must be workload-adaptive to prevent cognitive overload, and speech input must undergo rigorous semantic validation to prevent unsafe command execution. The architecture proposed in this paper formalizes this bidirectional regulation.

\subsection{Contextual and Cultural Variance in ASR}

The operational reliability of voice-driven interfaces depends on the transcription accuracy of the underlying Automatic Speech Recognition (ASR) engine. In multi-UAV supervisory control, ASR performance degrades due to two primary variables: environmental context and cultural phonetic variance. 

Regarding environmental context, outdoor UAV operations subject the microphone to severe, non-stationary acoustic masking from wind and rotor noise. Park and Na \cite{park2020front} demonstrated that such non-stationary noise drastically reduces trigger-word detection and command recognition in multi-UAV environments. While signal processing techniques like spectral subtraction attempt to filter this noise, they frequently induce signal distortion that further degrades transcription accuracy \cite{park2020front}.

Beyond acoustic noise, ASR exhibits severe performance disparities across cultural and phonetic demographics. Koenecke et al. \cite{koenecke2020racial} established that commercial ASR systems generate significantly higher Word Error Rates (WER) for non-standard dialects and minority demographics compared to standard phonetic profiles. These disparities directly impact human-robot interaction. Bingol and Aydogmus \cite{bingol2020performing} demonstrated that regional dialect variations degrade robotic speech recognition, necessitating custom acoustic modeling to handle pronunciation discrepancies. Similarly, Korayem et al. \cite{korayem2021design} showed that robotic voice command systems trained on native datasets experience significant accuracy drops when operated by non-native speakers. In a supervisory control environment, this cross-cultural bias means that non-native speakers or operators with distinct regional accents face higher rates of command misclassification.

The compounding effect of acoustic masking and cross-cultural phonetic variation renders raw ASR output unreliable for critical multi-agent task allocation. To guarantee system dependability, the interaction pipeline must include a semantic reasoning layer capable of mapping noisy, dialect-heavy transcriptions to exact, executable state constraints.

\subsection{Language Models in Multi-Agent Systems}
Large Language Models (LLMs) currently perform high-level task allocation and planning in robotic systems \cite{ahn2022can, vemprala2024chatgpt}. These architectures process natural language commands and generate sequential robot actions. However, direct LLM-to-robot control introduces significant operational risks in multi-agent environments. Cloud-based LLMs exhibit high computational latency, which violates the real-time constraints of swarm supervision. Furthermore, probabilistic text generation causes hallucinations, leading to non-deterministic and potentially catastrophic physical execution \cite{vemprala2024chatgpt}.

Our proposed architecture separates semantic interpretation from control logic. Direct end-to-end AI planners struggle to scale with the combinatorial complexity of multi-agent routing, often requiring human-in-the-loop intervention to escape local minima \cite{ameri2023interplay}. Therefore, instead of generating direct flight commands, the system employs a local Small Language Model (SLM) strictly as a semantic parser. The SLM translates transcribed speech into a structured format (JSON). This structured output defines exact operational constraints and target identifiers. Downstream, these constraints are processed by deterministic multi-drone planning algorithms, such as cooperative task allocation and behavior tree execution models \cite{ricardo2023cooperative}. This separation isolates the stochastic uncertainty of language processing from the safety-critical execution layer, ensuring reliable task generation under bounded latency.

\section{Speech Input}
\label{sec:system_overview1}

\label{sec:SpeechInputPolicy}

Speech input supports higher-level supervisory behaviors, including contextual references and summary queries that reduce cognitive load and improve situational awareness. However, it introduces challenges such as ambiguity, recognition uncertainty, and error handling, particularly in safety-critical scenarios where mistakes carry real consequences \cite{zhang2018deep}. These issues are further compounded by accent variability and differences in speech patterns, which can increase misinterpretation if not addressed. 

Speech input design must therefore balance efficiency with robustness by incorporating validation, confirmation mechanisms, and clear feedback. In practice, speech works best not as a replacement but as part of a multimodal system that enhances attention management, interaction speed, and overall operational control \cite{feng2025effects}.

\subsection{Speech Input Policy}
\label{sec:SpeechInputPolicy}

Speech input provides an efficient interaction modality for supervisory control tasks, enabling operators to issue commands without diverting visual attention. However, speech recognition introduces uncertainty, ambiguity, and potential recognition errors. To ensure robust interaction, we define a regulated speech input policy governing command interpretation, validation, and execution. This model is also presented in Algorithm \ref{alg:speechInput}.

\subsubsection{Command Representation}
To prevent referential ambiguity and enforce deterministic downstream execution, the STP outputs a strict JSON payload. The recognized command $c$ is formalized as:
\begin{equation}
    c = (i, t, p, k)
\end{equation}
where:
\begin{itemize}
    \item $i \in I$ represents the inferred operational intent. The intent space $I$ is a closed set of 13 valid actions (e.g., \texttt{INSPECT}, \texttt{SURVEY}, \texttt{ABORT}, \texttt{STATUS}). Unclear intents strictly map to a fallback \texttt{UNKNOWN} state.
    \item $t \in T$ identifies the addressed drone or team. The target space $T$ is restricted to predefined entity identifiers (e.g., \texttt{drone one}, \texttt{team Red}). Spatial locations are explicitly excluded from the target space.
    \item $p$ contains an optional array of extracted task parameters (e.g., specific target locations, required camera equipment).
    \item $k \in [0, 1]$ denotes the model's confidence scalar regarding the semantic interpretation.
\end{itemize}

\subsubsection{Recognition Constraints}
Speech interaction is guided by preferred grammatical patterns to reduce ambiguity, rather than strictly enforcing predefined command structures. This allows for flexibility in how users express commands, while verifying that only inputs with a correctly interpreted intent structure are accepted:

\begin{equation}
intent(c) \notin I_{allowed} \Rightarrow reject(c)
\end{equation}
where $I_{allowed}$ is a subset of $I$ which represents the set of intents that the target can perform given its current condition. For example, a drone which is not flying cannot execute a landing intent.

\subsubsection{Confidence-Based Validation}

Command execution is conditioned on recognition confidence using an execution threshold ($\theta_{exec}$) and a confirmation threshold ($\theta_{confirm}$):

\begin{equation}
confidence(c) \ge \theta_{exec} \Rightarrow execute(c)
\end{equation}

\begin{equation}
    \theta_{confirm} \le confidence(c) < \theta_{exec} \Rightarrow requestConfirmation(c)
\end{equation}

Low-confidence interpretations trigger clarification mechanisms rather than direct execution.

\subsubsection{Ambiguity Resolution}

Commands lacking a clear target or containing multiple possible interpretations invoke disambiguation procedures:

\begin{equation}
ambiguous(c) \Rightarrow initiateResolution(c)
\end{equation}

Resolution strategies may include visual highlighting, clarification prompts, or contextual inference.

\subsubsection{Error Recovery Policy}

Recognition errors are mitigated through explicit feedback and reversible actions. Critical commands (e.g., mission abort) require confirmation even at high confidence:

\begin{equation}
criticalIntent(c) \Rightarrow requireConfirmation(c)
\end{equation}

This mechanism intends to prevent catastrophic consequences of recognition failures.

\begin{algorithm}
\caption{Speech Input Policy}
\label{alg:speechInput}
\begin{algorithmic}[1]

\STATE \textbf{Procedure} OnSpeechRecognized($transcript$, $confidence$, $context$)

\STATE $c \gets ParseCommand(transcript, confidence)$
    
\IF{$c = \emptyset$}
    \STATE NotifyOperator(unrecognized\_utterance)
    \STATE \textbf{return}
\ENDIF

\IF{$intent(c) \notin I_{allowed}$}
    \STATE NotifyOperator(invalid\_intent)
    \STATE \textbf{return}
\ENDIF

\IF{$confidence(c) < \theta_{confirm}$}
    \STATE RequestRepeat()
    \STATE \textbf{return}
\ENDIF

\IF{$IntentRequiresTarget(c)$}
    \STATE $T \gets ResolveCandidates(c, context)$

    \IF{$|T| = 0$}
        \STATE NotifyOperator(no\_target)
        \STATE \textbf{return}
    \ENDIF

    \IF{$|T| > 1$}
        \STATE InitiateResolution($T$)
        \STATE \textbf{return}
    \ENDIF

    \STATE $target(c) \gets T[0]$
\ENDIF

\IF{$intent(c) \in I_{critical}$}
    \STATE RequireConfirmation($c$)
    \STATE \textbf{return}
\ENDIF

\IF{$confidence(c) \ge \theta_{exec}$}
    \STATE Execute($c$)
\ELSE
    \STATE RequestConfirmation($c$)
\ENDIF

\STATE \textbf{End Procedure}

\end{algorithmic}
\end{algorithm}

\section{Semantic Task Parser}
\label{sec:system_overview2}

\begin{figure}[ht!]
    \centering
    \includegraphics[width=0.7\columnwidth]{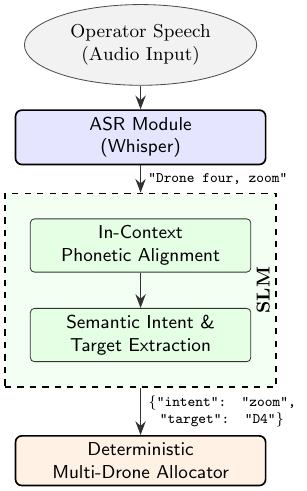}
    \caption{The system pipeline. Raw operator audio is transcribed by the ASR module. The SLM performs in-context phonetic alignment to resolve transcription errors and extracts structured state constraints (JSON) for downstream multi-agent task allocation.}
    \Description{Speech-command processing pipeline. Operator audio is transcribed using Whisper, then an SLM performs in-context phonetic alignment and semantic intent and target extraction. For the example ``Drone four, zoom,'' the SLM produces intent ``zoom'' and target ``D4'', which are passed to a deterministic multi-drone allocator.}
    \label{fig:pipeline}
\end{figure}

The system operates as a sequential semantic extraction pipeline, shown in Figure \ref{fig:pipeline}. An Automatic Speech Recognition (ASR) module first transcribes operator speech. A Small Language Model (SLM) performs joint semantic and phonetic inference on the raw transcripts to mitigate acoustic and dialectical distortions. The SLM resolves acoustically confusable terms using domain context and maps the validated utterance to a structured JSON payload defining exact intents and target identifiers. This decouples the uncertainty of speech recognition from the deterministic drone control.

\subsection{Semantic Task Parser Architecture}
\label{sec:SpeechInputPolicy}

Speech input in multi-agent environments amplifies referential ambiguity and recognition uncertainty. To ensure safe execution, the SLM operates as a regulated Semantic Task Parser (STP). The STP maps a raw natural language query ($q$) transcribed by the ASR module into a structured, executable task tuple, bypassing traditional rigid grammar constraints in favor of in-context probabilistic mapping.

The STP supports multi-intent sequential splitting. For compound queries, the SLM decomposes the utterance into a set of tuples (e.g., $c_1 = (\texttt{GOTO}, t, p_1, k_1)$ and $c_2 = (\texttt{INSPECT}, t, p_2, k_2)$). The complete operational state space bounding these parameters is defined in Table \ref{tab:state_space}.

\begin{table}[htbp]
\caption{Operational State Space for the Semantic Task Parser}
\label{tab:state_space}
\centering
\small
\begin{tabular}{@{}ll@{}}
\toprule
\textbf{Parameter} & \textbf{Permitted Vocabulary Space} \\ \midrule
\textbf{Intents ($I$)} & \texttt{INSPECT}, \texttt{SURVEY}, \texttt{SUSPEND}, \texttt{ABORT}, \\
& \texttt{STATUS}, \texttt{GOTO}, \texttt{LAND}, \texttt{ZOOM}, \texttt{VIDEO\_ON}, \\
& \texttt{VIDEO\_OFF}, \texttt{ALTITUDE}, \texttt{RTL}, \texttt{BATTERY} \\ \midrule
\textbf{Targets ($T$)} & \texttt{drone one}, \texttt{drone two}, \texttt{drone three}, \\
& \texttt{drone four}, \texttt{drone five}, \texttt{all drones}, \\
& \texttt{team red}, \texttt{team blue}, \texttt{team green} \\ \midrule
\textbf{Fallback} & \texttt{UNKNOWN} (Triggers $k < \theta_{exec}$ clarification) \\ \bottomrule
\end{tabular}
\end{table}

\subsubsection{In-Context Phonetic Alignment}
Raw ASR transcripts contain acoustic distortions. To mitigate these errors before semantic mapping, the STP executes a mandatory phonetic normalization process. The SLM identifies acoustic tokens that fail to match the closed vocabulary spaces $I$ or $T$. 

The STP applies structural context to guide this phonetic alignment. For example, the system expects a numerical identifier following the keyword "drone". If the ASR transcribes \texttt{drone too} or \texttt{drone tree}, the STP applies an internal phonetic confusion rule to map the tokens to \texttt{drone two} and \texttt{drone three}. The STP executes minimal token edits rather than full sentence rewrites to preserve the original operational syntax.

\subsubsection{Confidence-Based Validation Policy}
Command execution is conditioned on a prioritized decision sequence. The STP applies phonetic normalization, maps the tokens to valid intents ($i$) and targets ($t$), extracts parameters ($p$), and assigns the confidence scalar ($k$). 

The STP penalizes uncertain translations to ensure system safety. If the raw query $q$ required phonetic correction to achieve a valid state, the SLM reduces the output confidence $k$ by a minimum penalty of $0.1$. 
Commands with penalized confidence scalars that fall below the execution threshold $\theta_{exec}$ trigger clarification mechanisms rather than direct execution, preventing catastrophic consequences from recognition failures.

\section{Experimental Methodology}
\label{sec:experiments}


In this section, we present the implemented end-to-end pipeline to evaluate the proposed architecture. This methodology evaluates the system's resilience against ASR uncertainty and referential ambiguity across varied acoustic and cultural conditions.

\subsection{Pipeline Implementation and Experimental Baseline}

\begin{figure}[ht!]
    \centering
    \fbox{\includegraphics[width=0.7\columnwidth, trim=2.2cm 4cm 15.5cm 1.5cm, clip]{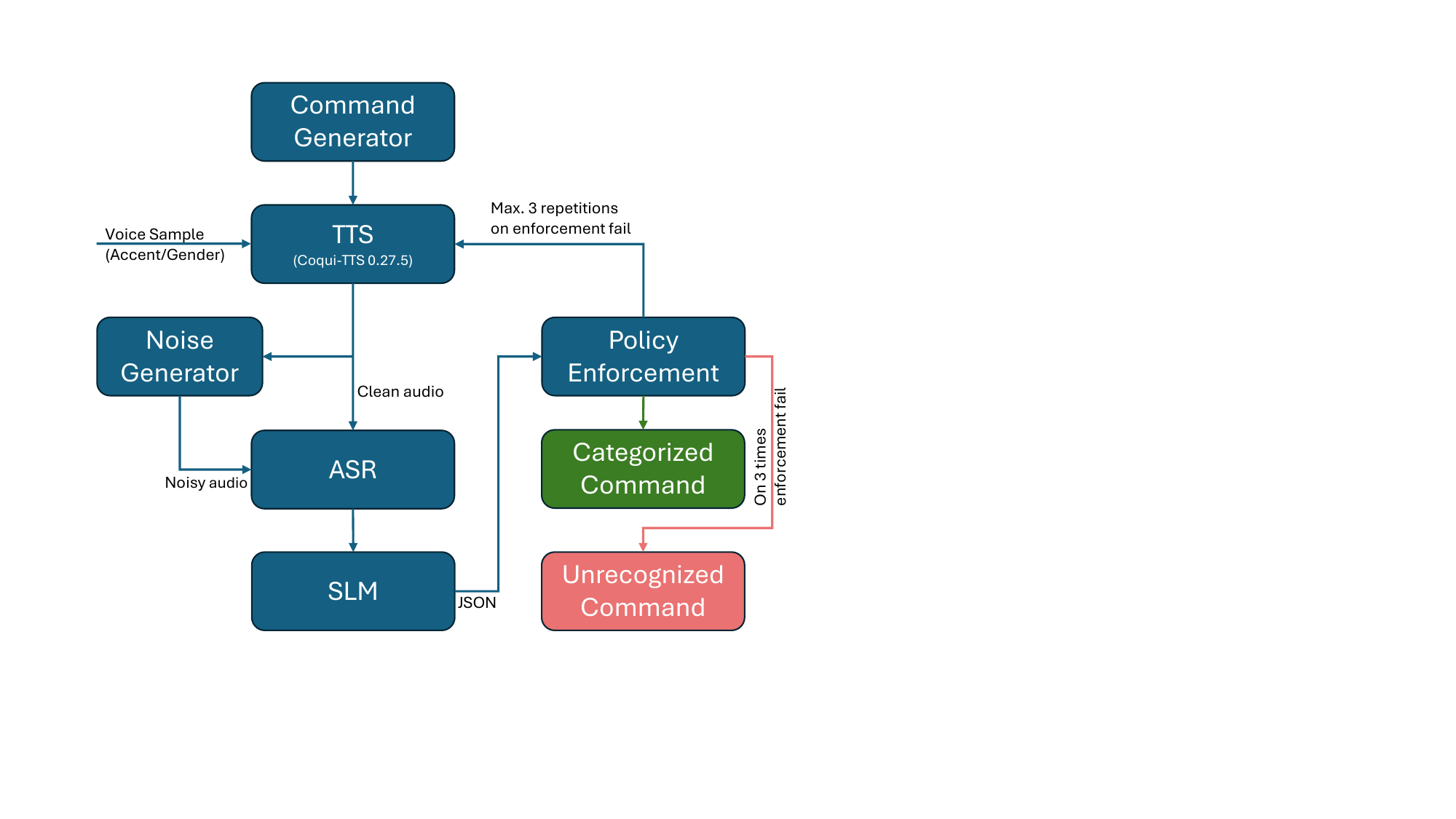}}
    \caption{Pipeline used for experiments. The output from TTS passes through the rest of the pipeline twice: (1) clean audio, (2) noisy audio.}
    \Description{ Command-processing pipeline used to generate and validate spoken commands. A command generator provides text to a text-to-speech system, which also receives a voice sample specifying accent and gender. The synthesized audio is sent directly to ASR as clean audio and is also passed through a noise generator to produce a noisy-audio variant for ASR. The ASR output is processed by an SLM, which produces a JSON representation that is sent to a policy-enforcement module. Recognized commands are categorized and accepted. Unrecognized commands are rejected, and failed enforcement can trigger up to three repetitions of the text-to-speech stage. }
    \label{fig:AppPipeline}
\end{figure}

The experimental pipeline (Figure \ref{fig:AppPipeline}) integrates three components: speech synthesis, speech transcription, and semantic parsing. Voice commands are generated using Coqui-TTS (version 0.27.5). This engine was selected due to its capability to synthesize speech conditioned on input sample audio files, enabling voice cloning and accent variations. The output is conditioned on 14 sample audio files \cite{speech_accent_archive} representing 11 distinct regional accents. For transcription, the Automatic Speech Recognition (ASR) module utilizes OpenAI's Whisper model. Finally, the Semantic Task Parser (STP) operates on Essential AI's Rnj1 SLM, executed locally within the LM Studio environment.

The SLM is evaluated under two distinct configurations to measure the impact of the proposed phonetic alignment layer:

\begin{itemize}
    \item \textbf{Without Phonetic Adjustment (Baseline):} The SLM operates as a standard semantic parser. It receives the raw ASR transcript and maps it to the JSON tuple $(i, t, p, k)$ based exclusively on standard natural language comprehension and few-shot examples. It contains no explicit rules for acoustic distortions.

    \item \textbf{With Phonetic Adjustment:} The SLM employs phonetic normalization rules. It actively executes joint semantic-phonetic inference, applies $0.1$ confidence penalty for modified tokens, and utilizes spatial and numerical context to resolve homophonic ASR errors.
\end{itemize}

\subsection{Stochastic Fault Injection}
Valid synthetic speech queries ($q$) were subjected to stochastic perturbation vectors to emulate realistic operational hazards in field deployments:
\begin{enumerate}
    \item \textbf{Acoustic Masking (Noisy Audio):} To simulate environmental noise (e.g., drone rotors, wind), background acoustic interference was injected into the synthesized TTS audio files prior to ASR processing. This directly degraded the transcription accuracy.
    \item \textbf{Referential Omission:} To model operator cognitive overload, the target parameter was truncated with probability $P_{omit} = 0.15$ (e.g., uttering "Return to base" instead of "Drone three, return to base"). This forces the SLM to rely on contextual inference or correctly assign the \texttt{UNKNOWN} target state.
    \item \textbf{Intent Corruption:} Benign intents were subjected to forced acoustic mutations, misclassifying them as critical actions with probability $P_{corrupt} = 0.05$ (e.g., misinterpreting "Report" as "Abort").
\end{enumerate}

\subsection{Dependability Metrics}

System dependability is quantified through objective metrics highlighting the engineering trade-off between safety and operational efficiency:
\begin{itemize}
    \item \textbf{Intent and Target Accuracy:} The percentage of correctly classified operational intents ($i$) and target identifiers ($t$) parsed from the generated JSON tuples.

    \item \textbf{Catastrophic Failure Rate (CFR):} The proportion of erroneously executed critical commands (e.g., mapping a distorted transcript to an unintended \texttt{ABORT} or \texttt{LAND} intent). The architectural goal is that $\text{CFR} \to 0$ irrespective of the noise variance $\sigma^2$, forcing highly noisy or uncertain inputs into clarification states rather than executing them.

    \item \textbf{Spurious Execution Rate (SER):} The rate at which non-critical, corrupted commands bypass the validation filter (e.g., incorrectly assigning a \texttt{STATUS} update to the wrong drone).
    
    \item \textbf{Intervention Request Rate (IRR):} The frequency of triggered clarification states ($k < \theta_{exec}$). This quantifies the operational cost of the fail-safe mechanism. A system with near-zero CFR but excessive IRR becomes operationally unusable due to constant operator prompting.
\end{itemize}

\section{Results and Discussion}
\label{sec:Discussion}

Table~\ref{tab:result_percentages} summarizes the performance of the pipeline in terms of intent and target classification accuracy under noisy and clean audio conditions, comparing results with and without phonetic adjustment. 

The results show a consistent improvement in performance when phonetic adjustment is applied to the pipeline. This effect is particularly pronounced under noisy audio conditions, where intent and target classification accuracies increase from 76.01\% to 84.64\%, and from 51.27\% to 70.43\% respectively. The combined intent and target accuracy also rises substantially, from 42.47\% to 62.85\%, indicating that phonetic adjustment significantly enhances robustness in challenging acoustic environments.

In clean audio conditions, performance is already relatively high, yet phonetic adjustment still improves the results. Intent accuracy increases from 84.84\% to 91.50\%, while target accuracy improves from 57.43\% to 75.41\%. The joint accuracy metric benefits as well, increasing from 50.56\% to 70.48\%. Even in clean audio, the baseline target accuracy remains relatively low (57.43\%) because standard ASR models frequently mistranscribe target identifiers, specifically numbers and colors into homophones (e.g., 'too' instead of 'two', or 'read' instead of 'red'). Illustrating that the unadjusted SLM fails to map the input to valid system entities. These improvements suggest that phonetic adjustment assists classification in noise-related degraded audios and also enhances overall model discrimination even in favorable conditions.

Overall, the results indicate that phonetic adjustment provides a substantial and consistent benefit, with the largest relative improvements observed in target and joint classification metrics, particularly in noisy settings. This suggests that phonetic-level normalization plays a critical role in improving model robustness and generalization.

\begin{table}[t]
\centering
\small
\setlength{\tabcolsep}{4pt}
\begin{tabular}{lcccccc}
\toprule
& \multicolumn{3}{c}{No Phonetic Adj.} 
& \multicolumn{3}{c}{With Phonetic Adj.} \\
\cmidrule(lr){2-4} \cmidrule(lr){5-7}
Cond. 
& Intent & Target & Int. and Tgt. 
& Intent & Target & Int. and Tgt. \\
\midrule
Noisy & 76.01 & 51.27 & 42.47 & 84.64 & 70.43 & 62.85 \\
Clean & 84.84 & 57.43 & 50.56 & 91.50 & 75.41 & 70.48 \\
\bottomrule
\end{tabular}
\caption{Percentages of correct intent and target classification in noisy and clean audio across all accents with and without phonetic adjustment after 10000 iterations.}
\label{tab:result_percentages}
\end{table}

Table \ref{tab:result_dep_metrics} summarizes the dependability performance of the system under noisy and clean conditions, comparing the results obtained with and without phonetic adjustment. The metrics reported include Catastrophic Failure Rate (CFR), Spurious Execution Rate (SER), and Intervention Request Rate (IRR). Across both conditions, the introduction of phonetic adjustment consistently reduces failure- and error-related rates, with the most pronounced improvements observed in noisy environments. These results indicate that phonetic adjustment plays a significant role in improving system reliability and robustness, particularly when operating under adverse acoustic conditions.

\begin{table}[t]
\centering
\small
\setlength{\tabcolsep}{4pt}
\begin{tabular}{lcccccc}
\toprule
& \multicolumn{3}{c}{No Phonetic Adj.} 
& \multicolumn{3}{c}{With Phonetic Adj.} \\
\cmidrule(lr){2-4} \cmidrule(lr){5-7}
Condition
& CFR & SER & IRR  
& CFR & SER & IRR   \\
\midrule
Noisy & 0.414 & 0.222 & - & 0.153 & 0.154 & - \\
Clean & 0.320 & 0.134 & 0.091 & 0.074 & 0.086 & 0.135 \\
\bottomrule
\end{tabular}
\caption{Dependability metrics.}
\label{tab:result_dep_metrics}
\end{table}


\subsection{Intent Classification}

\begin{figure}
    \centering
    \fbox{\includegraphics[width=0.96\linewidth, clip]{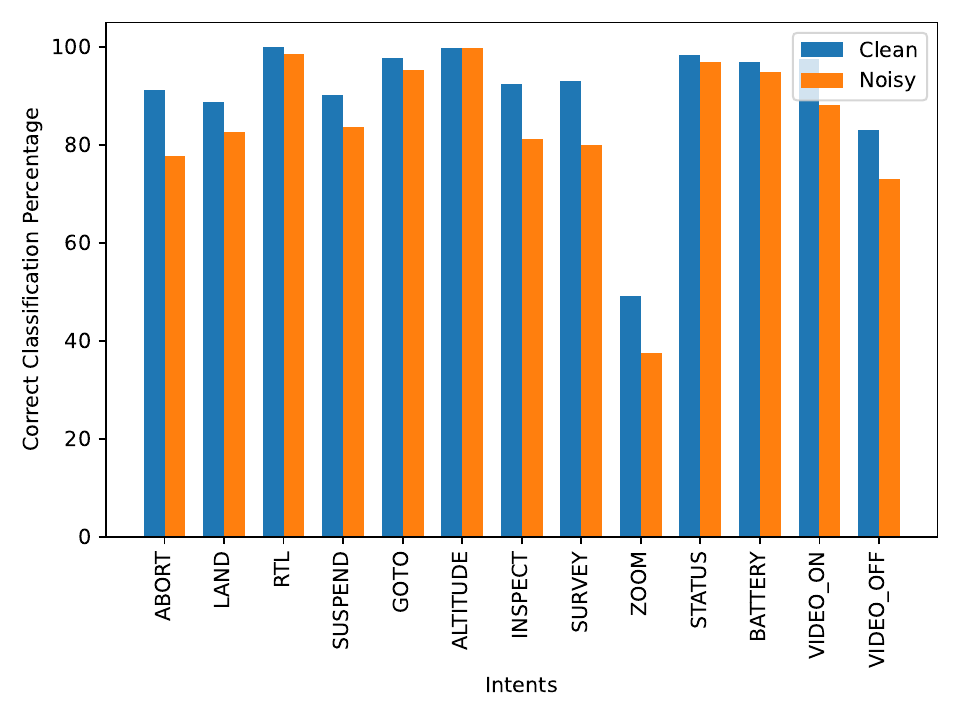}}
    \caption{Result's breakdown by intent with phonetic adjustment.}
    \Description{Grouped bar chart showing correct intent-classification percentages under clean and noisy audio conditions for 13 command intents. Clean-audio performance is generally high, mostly above 85 percent, while noisy-audio performance is consistently lower for most intents. RTL, ALTITUDE, STATUS, and BATTERY remain near or above 95 percent in both conditions. ZOOM has the lowest accuracy, dropping from about 49 percent in clean audio to about 37 percent with noise. Larger noise-related decreases are also visible for ABORT, INSPECT, SURVEY, and VIDEO\_OFF.
}
    \label{fig:IntentBreakdownP3}
\end{figure}

To provide a more granular understanding of the system’s performance, Figure \ref{fig:IntentBreakdownP3} presents intent-wise classification accuracy in clean and noisy conditions with phonetic adjustment. A prominent observation is the significantly lower performance associated with "zoom" intent compared to all other categories. Detailed inspection of debug logs suggests that this degradation is a result of two primary factors: First, the pronunciation of “zoom” exhibits substantial variability across different accents, increasing acoustic ambiguity. Second, the term is phonetically similar to a range of common words (e.g., “soon,” “zone,” “seen,” “person” \footnote{In this case the complete sentence was "Drone four, zoon". The ASR transcribed it as "Drone person" concatenating "four" and "zoom".}), leading to frequent misrecognitions at the ASR level. Indeed, multiple instances were observed where “zoom” was transcribed as these acoustically analogous alternatives, directly impacting downstream intent classification.

More broadly, the results indicate that noisy conditions consistently degrade classification performance in all intents. However, this degradation is not uniform. Intents with simpler or shorter phonetic structures, such as "zoom" and "survey", are disproportionately affected, likely due to increased susceptibility to noise-induced ambiguity and lexical confusion. In contrast, intents such as "altitude" demonstrate near-perfect classification accuracy even under noisy conditions. This can be attributed to their greater phonetic complexity and distinctiveness, which reduce overlap with other lexical items and make them more robust to both noise and ASR errors.

Overall, these findings highlight that intent classification performance is strongly influenced not only by acoustic conditions but also by the intrinsic phonetic properties of the target vocabulary. Intents with higher lexical uniqueness and phonetic richness appear inherently more resilient, whereas shorter, acoustically confusable terms remain a critical challenge for robust spoken language understanding systems.

\subsection{Accents and Gender}
The experiments are conducted using 14 distinct voice samples representing 11 different accents, providing a diverse evaluation of system performance across varied speech patterns. The male voice set includes speakers from US Arkansas (ARK), Australia (AUS), Canada (CAN), India (IND), Ireland (IRL), New Zealand (NZL), South Africa (SAF), and US Virginia (USV). The female voice set includes speakers from Canada (CAN), Britain (ENG), India (IND), Jamaica (JAM), Malaysia (MAL), and New Zealand (NZL). This distribution ensures coverage of a broad range of regional pronunciations and phonetic variations, allowing a more comprehensive assessment of the system’s robustness to accent diversity.

\begin{figure}
    \centering
    \fbox{\includegraphics[width=0.96\linewidth, clip]{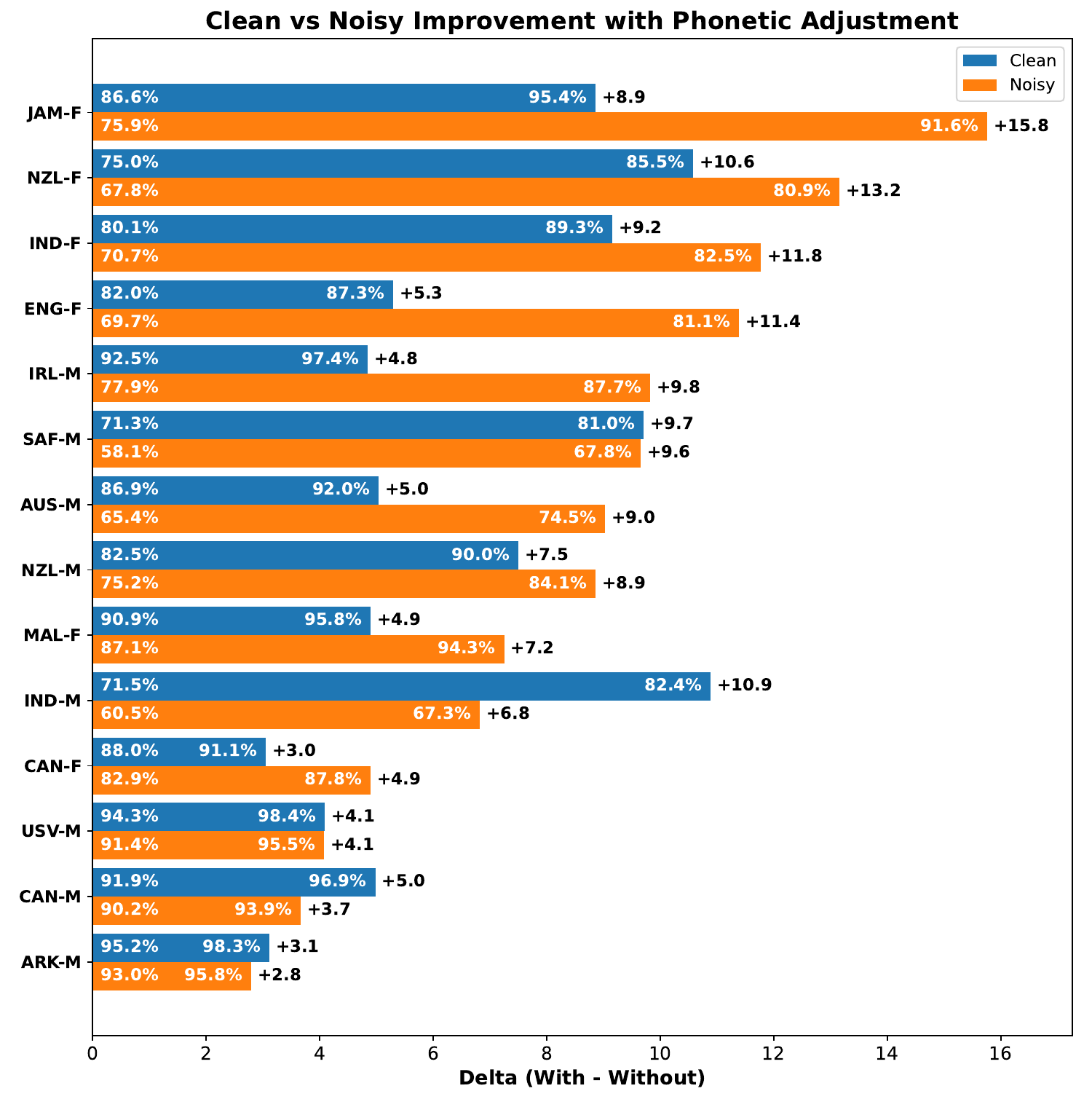}}
    \caption{Improvements in intent classification on different accents in clean and noisy audio.}
    \Description{Classification-accuracy improvements from phonetic adjustment across accent and gender groups under clean and noisy audio. Every group improves after adjustment, with larger gains generally observed in noisy conditions. The strongest noisy-condition gains occur for JAM-F and NZL-F, while clean audio improvements are more modest. Groups with already high baseline accuracy, including ARK-M and USV-M, show the smallest improvements.}
    \label{fig:AccentsDiff}
\end{figure}

Figure \ref{fig:AccentsDiff} presents the impact of phonetic adjustment on classification performance in the aforementioned accent groups under both clean and noisy conditions. A consistent improvement is observed across all accents, with gains in clean speech remaining relatively modest and tightly distributed, generally within a range of approximately +3 to +11 percentage points. In contrast, the improvements in noisy conditions are larger and more variable, frequently exceeding +10 points and reaching as high as +15.8 for Jamaican English. This indicates that while phonetic adjustment provides incremental refinement in ideal acoustic settings, its contribution becomes substantially more marked when signal degradation is present.

A clear pattern also emerges: Accents with lower baseline performance, such as Indian and South African English, exhibit comparatively larger gains, particularly in noisy environments. 
Conversely, accents with already high baseline performance, including American and Canadian English, show smaller but still consistent improvements, reflecting diminishing returns in near-ceiling conditions while confirming that the method provides generalizable benefits across accent groups.


An additional clear observation is that phonetic adjustment yields disproportionately larger gains for female voices under noisy conditions. Specifically, the four largest improvements observed in the noisy setting all correspond to female speakers. This trend becomes even clearer when comparing gender pairs with the same accent. For example, in the New Zealand subset, the female voice (NZL-F) shows a performance increase of 13.2\%, compared to 8.9\% for the male counterpart (NZL-M). A similar pattern appears for Indian accents, where IND-F improves by 11.8\%, while IND-M shows a more modest improvement of 6.8\%.

This disparity is especially interesting given that the baseline classification accuracy for IND-F is already higher than that of IND-M (70.7\% versus 60.5\%). Even with this higher starting point, the relative improvement for the female voice remains substantially larger, suggesting that phonetic adjustment may be especially effective for female speech characteristics in degraded acoustic environments.

More broadly, the results indicate that while phonetic adjustment consistently improves performance across all speaker groups, its impact is generally larger in noisy conditions and appears to interact with gender in a non-uniform manner. This may point to underlying differences in how noise affects phonetic features in male versus female speech, or how the adjustment mechanism compensates for these distortions.

Overall, the results highlight the role of phonetic adjustment as a robustness-enhancing mechanism rather than merely a fine-tuning strategy. Its disproportionate impact under noisy conditions underscores its capacity to stabilize performance in the presence of acoustic variability, particularly for underrepresented or phonetically distinct accents.



\subsection{Discussion}

\subsubsection{Poor TTS Output}
A detailed inspection of the application logs revealed that some misclassifications originate from degraded TTS output, in certain cases rendering the audio difficult to understand even for human listeners. While this could be considered a complicating factor, it is representative of real-world conditions, where speech input is often affected by factors such as articulation variability, background noise, or concurrent activities (e.g., eating or drinking). In this context, imperfect TTS output provides a more rigorous evaluation of the processing pipeline than idealized, high-quality speech.

\subsubsection{Classification Ambiguity}
Classification errors can also occur when different commands have similar meanings in natural language. For example, the system struggles to distinguish between the intents "abort" and "suspend". In everyday speech, both terms indicate stopping an action. In drone control, they execute entirely different behaviors. "Abort" terminates the mission and forces the drone to return or land. "Suspend" pauses the current task and commands the drone to hover in place. 

When operators use phrases like "stop what you are doing" or "halt", the SLM may map the speech to the wrong intent. To resolve this overlap, the SLM requires exact task definitions and few-shot examples in its prompt. This ensures the model maps ambiguous human speech to the correct operational state.

\subsubsection{Acoustic and Phonetic Ambiguity}

ASR transcription errors occur mainly due to phonetic similarity, different accents, and background noise. Analyzing the system logs shows specific error patterns that cause standard string-matching and rigid grammar rules to fail. First, the ASR frequently confuses numbers and colors with words that sound the same (homophones). For example, it transcribes "one" as "won", "two" as "to" or "too", "four" as "for" or "fore", "blue" as "blew", and "red" as "read". 

Second, varying accents and background noise break down specific technical words. The target word "drone" has many variations in the transcriptions, appearing as "drawn", "drown", "drownd", "roan", "room", "drum", or "draw". Similarly, the command "zoom" appears as "zum", "zam", "zom", "som", or "some". Lastly, the ASR often splits compound words into unrelated text. It transcribes "waypoint" as "weightpoint", "weight point", or "eight point". Furthermore, the ASR changes action commands into words that sound similar but make no sense in this context. For example, it transcribes "show feed" as "Sophie", "show" as "shoal", and "drone two" as "drone tool".

These failures illustrate why the SLM phonetic adjustment is necessary, it evaluates the text based on the context of drone operations. Therefore, it maps a wrong transcription like "Sophie" back to the valid command "show feed". This contextual correction recovers executable commands from failed ASR transcriptions by interpreting the intended meaning within the drone context, as shown in our results.

\section{Conclusion and Future Work}
\label{sec:Conclusion}

This paper presented a speech interaction framework for supervising multiple UAVs, designed to remain reliable under noise and accent variation. The approach combines a Small Language Model (SLM) with a regulated input policy to interpret speech and validate commands before execution, reducing the risks associated with ASR errors. The SLM is also instructed to take phonetic variations into consideration while interpreting the commands. The results show that phonetic adjustment improves both accuracy and system reliability, especially in noisy conditions. In particular, the system achieves better intent and target recognition while significantly lowering critical failure rates. Overall, these findings demonstrate that structured and validated speech interaction can serve as a practical and safe complement to traditional control methods in multi-UAV systems.

Future work should focus on evaluating the system with real users in operational environments, where factors such as spontaneous speech and unpredictable noise may introduce additional challenges. Incorporating adaptive mechanisms could allow the system to better handle individual speaking styles and accents over time. Another promising direction is integrating speech with other modalities, such as visual context or gestures, to improve disambiguation and reduce the need for operator intervention. Finally, refining uncertainty handling and confidence thresholds could help balance safety and efficiency by minimizing unnecessary confirmations while still preventing unsafe actions.

\section*{Safe and Responsible Innovation}
This work addresses safety-critical human–multi-UAV interaction, where misinterpretation of speech commands can lead to operational risks. To mitigate this, the proposed architecture incorporates confidence-based validation, ambiguity resolution, and mandatory confirmation for critical actions, reducing the likelihood of unsafe execution. The system explicitly accounts for bias in Automatic Speech Recognition (ASR) by evaluating performance across diverse accents and conditions, promoting inclusivity. Synthetic data is used to avoid privacy concerns associated with real user recordings. The framework is also designed to prioritize human oversight, transparency, and fail-safe behavior in deployment.

\vspace{-1mm}
\section*{acknowledgment}
\vspace{-1mm}

This research was funded by the Horizon Europe research and innovation program of the European Union and the Chips Joint Venture under GA No. 101194287, NexTArc (Next Generation Open Innovations in Trustworthy Embedded AI Architectures for Smart Cities, Mobility, and Logistics).

\bibliographystyle{unsrt}  
\bibliography{references}  

\end{document}